\documentstyle[preprint,eqsecnum,epsfig,aps]{revtex}

\begin{document}
\title{Charge exchange in relativistic heavy ion collisions}
\author{C. A. Bertulani \thanks{%
e-mail: bertu@if.ufrj.br} and D. S. Dolci \thanks{%
e-mail: dolci@if.ufrj.br}}
\address{{\small Instituto de F\'\i sica, Universidade Federal do Rio de Janeiro}\\
{\small 21945-970 \ Rio de Janeiro, RJ, Brazil.}}
\maketitle

\begin{abstract}
Elastic charge-exchange in relativistic heavy ion collisions is responsible for the
non-disruptive change of the charge state of the nuclei. We show that it 
can be reliably calculated within the eikonal approximation for the 
reaction part.  The
formalism is applied to the charge-pickup cross sections of
158 GeV/nucleon Pb projectiles on several targets. The relative
contributions of pion- and rho-exchange are determined, using 
a single-particle model for the internal structure of the nuclei. 
The calculated cross sections are non-negligible for heavy
targets.
It is shown that these cross sections can be useful to obtain information
on the Gamow-Teller transition strengths of the nuclei.
\end{abstract}


\begin{title}
{\bf Charge Exchange in \\
Relativistic Heavy Ion Collisions \rm}
\end{title}

\vspace{0.5in}

\section{\bf Introduction}

Heavy-ion charge-pickup reactions can help us to
access useful information on beta-decay transition strengths. At low energy
nuclear collisions, it is well known that charge-exchange is accomplished by
the proton exchange between the nuclei \cite{feshbach}. At intermediate
energies ( $\sim $100 MeV/nucleon) nucleon-exchange competes with the
charged meson exchange \cite{lenske}. At relativistic energies ( $>$ 1
GeV/nucleon), the charge stripping, $\Delta Z=-1$, cross sections will have a
substantial contribution from proton removal \cite{CH92}, but the charge pickup, $\Delta
Z=+1$,  cross section will be solely due to the charged meson
exchange, mainly $\pi ^{\pm }$- and $\rho ^{\pm }$-exchange \cite{Ber93}. It
is virtually impossible that a proton-pickup will occur at these energies.
The strong absorption of heavy ions selects large impact parameters and
therefore emphasizes the longest range part of the charge exchange force.
Since the reaction is very peripheral, one expects that the charge-exchange
process is practically determined by the participation of the valence
nucleons. Therefore, these reactions should be a probe of charge
exchange in a nuclear environment. A series of studies of this type has
been performed theoretically and experimentally \cite{Oset:1989cd,Udagawa:1994ym,Ramstein:1999hc,Sneppen:1994zx}.

The charge-pickup cross sections for relativistic heavy ion collisions
have recently been studied at CERN \cite{hir95,gee95,dfub98,scheid}. 
Theoretically, little is
known about these reactions. It
is the aim of this paper to develop a simple description of charge-pickup
reactions in relativistic heavy ion collisions in terms of
what we believe to be the most important ingredients, namely the microscopic 
$\pi $- and $\rho $-exchange potentials. An eikonal approach developed in
ref. \cite{Ber93} for charge-exchange in nucleus-nucleus scattering is used.
Simple expressions are found which can be useful for estimation purposes in
the planning of future experiments. 

One basic assumption of our work is that the dominant process
is pion and rho exchange between a projectile and a target nucleon.
However, other mechanisms could also be of equal importance, 
i.e. $N+N \longrightarrow N + \Delta$
followed by absorption. There has been a quite large number of
measurements on spin-isospin excitations in charge-exchange (
$^3He,\, t$) experiments at
SATURNE that show the  physics of these kind of processes are
quite complicated. We will not study these questions in the present 
paper, leaving this for a future work. But, we refer to some of the SATURNE
papers \cite{Ell83,Ell85,Con86,Ber87} for the interested reader.

In section 2 we describe the reaction
mechanism and the structure ingredients of our calculation. In section 3 the
dependence of the cross sections on the number of participants 
surface nucleons is
studied in details and an application is done for the charge-pickup
cross sections for 158 GeV/nucleon Pb projectiles on several
targets. In section 4 the proportionality of the measured cross sections
to the Gamow-Teller strengths is shown. A comparison to the results
of section 3 shows that the reaction mechanism does not dependent on the
structure model.
In section 5 we present our conclusions.

\section{\bf Relativistic heavy ion charge-exchange reactions}

Following ref. \cite{Ber93}, the total cross section for 
charge-pickup cross sections in high energy heavy ion collisions is given
by 
\begin{equation}
\sigma =2\pi \int_{0}^{\infty }b\,{\cal P}(b)\,db\,,  \label{cross}
\end{equation}
where ${\cal P}$ is the probability of one-boson-exchange at the impact
parameter $b$ and is given by 
\begin{equation}
{\cal P}(b)=\biggl( {\frac{1}{4\pi ^{2}\hbar v}}\biggr)^{2}%
\,(2j_{P}+1)^{-1}(2j_{T}+1)^{-1}\;\exp \bigl\{-2\,Im\;\chi (b)\bigr\}%
\,\sum_{\nu ,\,{\bf m}}\,\biggl\vert M({\bf m},\,\nu ,\,b)\biggr\vert^{2}\,,
\label{prob}
\end{equation}
where $Im\;\chi (b)$ is the imaginary part of the eikonal phase. In high
energy collisions the phase $\chi (b)$ will be predominantly imaginary and
can be constructed from the $t$-matrix for nucleon-nucleon scattering \cite
{HRB91}.

The matrix element $M({\bf m},\,\nu ,\,b)$ carries information on the
nuclear structure, and is given by 
\begin{equation}
M({\bf m},\,\nu ,\,b)=\int_{0}^{\infty }dq\,q\,J_{\nu }(qb)\int_{0}^{2\pi
}\,d\phi _{q}\,e^{-i\nu \phi _{q}}\;{\cal M}({\bf m},\,{\bf q})\;,
\label{m2}
\end{equation}
with 
\begin{equation}
{\cal M}({\bf m},\,{\bf q})=<\Phi _{f}^{P}({\bf r}_{P})\,\Phi _{f}^{T}({\bf r%
}_{T})|e^{-i{\bf q.r}_{P}}\,V({\bf q})\,e^{i{\bf q.r}_{T}}|\Phi _{i}^{P}(%
{\bf r}_{P})\,\Phi _{i}^{T}({\bf r}_{T})>\, ,  \label{m1}
\end{equation}
where $\Phi^{P(T)}_{f(i)}$ are the internal wavefunctions of the projectile (P) and target (T) in the initial (i) and final (f) states, respectively.

In eq. (\ref{prob}), ${\bf m}=(m_{T},\,m_{T}^{\prime
},\,m_{P},\,m_{P}^{\prime })$ is the set of angular momentum quantum numbers
of the projectile and target wavefunction. ${\bf m}$ is measured along the
beam axis, and the sub indexes $T$ and $P$ refer to the target and
projectile, respectively. The probability is obtained by an average of
initial spins and a sum over final spins ($m_{P(T)};\,m_{P(T)}^{\prime }$).
The interaction potential responsible for the charge-exchange between the
nuclei is given by $V({\bf q})$, where ${\bf q}$ is the Fourier transform of
the microscopic charge-exchange interaction. For more details, see \cite
{Ber93}.

The equations (\ref{cross}) and (\ref{prob}) are the basic results of the
eikonal approach to the description of heavy-ion charge-exchange reactions
at relativistic energies. They can also be used for the calculation of the
excitation of $\Delta $ particles in nucleus-nucleus peripheral collisions.
The essential quantity to proceed further is the matrix element given by eq.
(\ref{m1}) which is needed to calculate the impact-parameter-dependent
amplitude $M({\bf m},\,\nu ,\,b)$ through eq. (\ref{m2}). The magnitude of
this amplitude decreases with the decreasing overlap between the nuclei,
i.e. with the impact parameter $b$. At small impact parameters the strong
absorption will reduce the charge-exchange probability. Therefore, we expect
that the probability given by eq. (\ref{prob}) is peaked at the grazing
impact parameter.

The pion+rho exchange potential, modified so that the zero range force is
corrected for an extended source function \cite{BE87}, can be written as 
\begin{equation}
V({\bf q})=V_{\pi }({\bf q})+V_{\rho }({\bf q})=\biggl[ v({\bf q})(\mathbf {\sigma}
_{1}\cdot \hat{{\bf q}})(\mathbf{\sigma} _{2}\cdot \hat{{\bf q}})+w({\bf q})\,(\mathbf{\sigma}
_{1}\cdot \mathbf{\sigma} _{2})\biggr] \,(\mathbf{\tau} _{1}\cdot \mathbf{\tau} _{2})\,,  \label{(F.5)}
\end{equation}
where 
\begin{equation}
v({\bf q})=v_{\pi }^{tens}({\bf q})+v_{\rho }^{tens}({\bf q})\,,\;\;\;\;\;\;%
\text{and}\;\;\;\;w({\bf q})=w_{\pi }^{cent}({\bf q})+\xi \,w_{\rho }^{cent}(%
{\bf q})+w_{\pi }^{tens}({\bf q})+w_{\rho }^{tens}({\bf q})\,,
\label{(F.6a)}
\end{equation}
and 
\begin{equation}
v_{\pi }^{tens}({\bf q})=-J_{\pi }\,{\frac{{\bf q}^{2}}{m_{\pi }^{2}+{\bf q}%
^{2}}}\,,\qquad v_{\rho }^{tens}({\bf q})=J_{\rho }\,{\frac{{\bf q}^{2}}{%
m_{\rho }^{2}+{\bf q}^{2}}}\,,  \label{(F.7b)}
\end{equation}

\begin{equation}
w_{\pi }^{cent}({\bf q})=-{\frac{1}{3}}\,J_{\pi }\,\biggl[ {\frac{{\bf q}^{2}%
}{m_{\pi }^{2}+{\bf q}^{2}}}-3\,g_{\pi }^{\prime }\biggr]\,,\qquad
\;\;\;w_{\rho }^{cent}({\bf q})=-{\frac{2}{3}}\,J_{\rho }\,\biggl[ {\frac{%
{\bf q}^{2}}{m_{\rho }^{2}+{\bf q}^{2}}}-{\frac{3}{2}}\,g_{\rho }^{\prime }%
\biggr]\,,  \label{(F.7d)}
\end{equation}

\begin{equation}
w_{\pi }^{tens}({\bf q})={\frac{1}{3}}\,J_{\pi }\,{\frac{{\bf q}^{2}}{m_{\pi
}^{2}+{\bf q}^{2}}}\,,\qquad w_{\rho }^{tens}({\bf q})=-{\frac{1}{3}}%
\,J_{\rho }\,{\frac{{\bf q}^{2}}{m_{\rho }^{2}+{\bf q}^{2}}}\,.
\label{(F.7f)}
\end{equation}
with the parameters $g_{\pi }^{\prime }=1/3$, $g_{\rho }^{\prime }=2/3$, and 
$\xi =0.4$ \cite{AB77}.

The values of the coupling constants $J_{\pi }$ and $J_{\rho }$ in nuclear
units are given by 
\begin{eqnarray}
J_{\pi } &=&{\frac{f_{\pi }^{2}}{m_{\pi }^{2}}}\equiv f_{\pi }^{2}\,{\frac{%
(\hbar c)^{3}}{(m_{\pi }c^{2})^{2}}}\simeq 400\;MeV\;fm^{3}  \nonumber \\
J_{\rho } &=&{\frac{f_{\rho }^{2}}{m_{\rho }^{2}}}\equiv f_{\rho }^{2}\,{%
\frac{(\hbar c)^{3}}{(m_{\rho }c^{2})^{2}}}\simeq 790\;MeV\;fm^{3}
\label{(F.8)}
\end{eqnarray}
where $f_{\pi }^{2}/4\pi =0.08$ ($f_{\rho }^{2}/4\pi =4.85$), $m_{\pi
}c^{2}=145$ MeV, and $m_{\rho }c^{2}=770$ MeV.

Turning off the terms $w^{cent}_{\pi , \rho}$, or $v^{tens}_{\pi , \rho}$
and $w^{tens}_{\pi , \rho}$, allows us to study the contributions from the
central and the tensor interaction, and from $\pi$- and $\rho$-exchange,
respectively.

Using eq. (\ref{(F.5)}), and single-particle wavefunctions, $\phi _{j\ell m}$%
, it is straightforward to show that eq. (\ref{m1}) reduces to \cite{Ber93}
(note that here the indexes $\pi$ and $\nu$ denote the proton and the
neutron wavefunctions, respectively) 
\begin{eqnarray}
&&{\cal M}({\bf m},\,{\bf q})=w({\bf q})\,\sum_{\mu }<\phi _{j_{T}^{\prime
}\ell _{T}^{\prime }m_{T}^{\prime }}^{(\pi )}|\sigma _{\mu }\,e^{i{\bf q.r}
}|\phi _{j_{T}\ell _{T}m_{T}}^{(\nu )}>\,<\phi _{j_{P}^{\prime }\ell
_{P}^{\prime }m_{P}^{\prime }}^{(\nu )}|\sigma _{\mu }\,e^{-i{\bf q.r}%
}|\phi _{j_{P}\ell _{P}m_{P}}^{(\pi )}>  \nonumber \\
&+&{\frac{4\pi }{3}}\,v({\bf q})\,\sum_{\mu \mu ^{\prime }}Y_{1\mu }(\hat{%
{\bf q}})\,Y_{1\mu ^{\prime }}(\hat{{\bf q}})\,<\phi _{j_{T}^{\prime }\ell
_{T}^{\prime }m_{T}^{\prime }}^{(\pi )}|\sigma _{\mu }\,e^{i{\bf q.r}}|\phi
_{j_{T}\ell _{T}m_{T}}^{(\nu )}>\,<\phi _{j_{P}^{\prime }\ell _{P}^{\prime
}m_{P}^{\prime }}^{(\nu )}|\sigma _{\mu ^{\prime }}\,e^{-i{\bf q.r}}|\phi
_{j_{P}\ell _{P}m_{P}}^{(\pi )}>  \nonumber \\
&&  \label{(B.5)}
\end{eqnarray}

Expanding $e^{i{\bf q.r}}$ into multipoles we can write 
\begin{equation}
<\phi _{j^{\prime }\ell ^{\prime }m^{\prime }}^{(\pi )}|\sigma _{\mu }\,e^{i%
{\bf q.r}}|\phi _{j\ell m}^{(\nu )}>=4\pi \sum_{IM}i^{I}\,Y_{IM}^{*}(\hat{%
{\bf q}})\,<\phi _{j^{\prime }\ell ^{\prime }m^{\prime }}^{(\pi
)}|j_{I}(qr)\,Y_{IM}(\hat{{\bf r}})\,\sigma _{\mu }|\phi _{j\ell m}^{(\nu
)}>\,.  \label{(B.6)}
\end{equation}
Since $j_{I}(qr)\,Y_{IM}(\hat{{\bf r}})$ is an irreducible tensor 
\begin{equation}
\sigma _{\mu }\;j_{I}(qr)\,Y_{IM}(\hat{{\bf r}})=\sum_{I^{\prime }M^{\prime
}}(I1M\mu |I^{\prime }M^{\prime })\,\Psi _{I^{\prime }M^{\prime }}\,,
\label{(B.7)}
\end{equation}
where $\Psi _{I^{\prime }M^{\prime }}$ is also an irreducible tensor.
Therefore, 
\begin{eqnarray}
&<&\phi _{j^{\prime }\ell ^{\prime }m^{\prime }}^{(\pi )}|\sigma _{\mu
}j_{I}(qr)\,Y_{IM}(\hat{{\bf r}})|\phi _{j\ell m}^{(\nu )}>=\sum_{I^{\prime
}M^{\prime }}(I1M\mu |I^{\prime }M^{\prime })\,<\phi _{j^{\prime }\ell
^{\prime }m^{\prime }}^{(\pi )}|\Psi _{I^{\prime }M^{\prime }}|\phi _{j\ell
m}^{(\nu )}>  \nonumber \\
&=&\sum_{I^{\prime }M^{\prime }}\;(I1M\mu |I^{\prime }M^{\prime
})\,(jI^{\prime }mM^{\prime }|j^{\prime }m^{\prime })\,<\phi _{j^{\prime
}}^{(\pi )}||\Psi _{I^{\prime }}||\phi _{j}^{(\nu )}>\,.  \label{(B.8)}
\end{eqnarray}

The eqs. (\ref{(B.5)}-\ref{(B.8)}) allows one to calculate the
charge-exchange between single-particle orbitals. The quantity needed is the
reduced matrix element $<\phi _{j^{\prime }}^{(\pi
)}\,||\,[j_{I}(qr)\,\sigma \otimes Y_{I}]_{I^{\prime }}\,||\,\phi _{j}^{(\nu
)}>$. These are calculated in textbooks of nuclear structure (see, e.g., 
\cite{La80}). If several orbitals contribute to the process, the respective
amplitudes can be added and further on averaged in the cross sections.

\section{\bf Single-particle matrix-elements and surface approximation}

The reduced matrix elements are calculated in the single-particle model for
neutrons and protons. Using eq. (A.2.24) of ref. \cite{La80} one finds 
\begin{eqnarray}
&<&\phi _{j^{\prime }}^{(\pi )}\,||\,[j_{I}(qr)\,\sigma \otimes
Y_{I}]_{I^{\prime }}\,||\,\phi _{j}^{(\nu )}>=-\left( -1\right) ^{\ell +\ell
^{\prime }+j^{\prime }-1/2}\left\{ \frac{2j+1}{4\pi (2I^{\prime }+1)}%
\right\} ^{1/2}\left( jj^{\prime }\frac{1}{2}-\frac{1}{2}\mid I^{\prime
}0\right)  \nonumber \\
&&\times \left\{ \frac{1+\left( -1\right) ^{\ell +\ell ^{\prime }+I}}{2}%
\right\} \,\left( 
\begin{array}{ll}
\sqrt{I+1} & 1/\sqrt{I+1} \\ 
0 & \sqrt{\left\{ \left( 2I+1\right) /I(I+1)\right\} } \\ 
-\sqrt{I} & 1/\sqrt{I}
\end{array}
\right) \left( 
\begin{array}{l}
1 \\ 
\left( -1\right) ^{\ell +1/2-j}\eta _{I^{\prime }}\left( jj^{\prime }\right)
\end{array}
\right) {\cal F}_{Ijj^{\prime }}\left( q\right)  \nonumber \\
&&  \label{(B.11a)}
\end{eqnarray}
where 
\begin{equation}
{\cal F}_{Ijj^{\prime }}\left( q\right) \,=\int_{0}^{\infty }R_{j}^{(\pi
)}(r)R_{j^{\prime }}^{(\nu )}(r)\,j_{I}(qr)\,r^{2}\,dr\,.  \label{(B.11b)}
\end{equation}
and

\begin{equation}
\eta _{I^{\prime }}\left( jj^{\prime }\right) =\frac{1}{2}\left\{ \left(
2j+1\right) +\left( -1\right) ^{j+j^{\prime }-I^{\prime }}\left( 2j^{\prime
}+1\right) \right\}   \label{(B.11c)}
\end{equation}
In eq. (\ref{(B.11b)}), $R_{j}^{(\pi )}(r)$ and $R_{j^{\prime }}^{(\nu )}(r)$
are the proton and neutron single particle radial wavefunctions,
respectively. 

The (3$\times $2) and (2$\times $1) arrays in the equation are matrices and
matrix multiplication is implied. The resulting matrix $A_{mn}$ is a (3$%
\times $1) array with $A_{11}$, $A_{21}$ and $A_{31}$ the values of the
reduced matrix elements when $I^{\prime }=I+1$, $I$, and $I-1,$
respectively. The factor $\left[ 1+\left( -1\right) ^{\ell +\ell ^{\prime
}+I}\right] /2$ vanishes unless $\ell +\ell ^{\prime }+I$ is even, that is
unless parity is conserved.

The proton and neutron wavefunctions which are needed for the integral eq. (%
\ref{(B.11b)}) must carefully account for the Pauli blocking in the final
state. Moreover, the pion, or $\rho ,$ is readily absorbed in the nuclear
surface and only the nucleons in the last shells and close to the nuclear
surface will contribute to the process. The surface approximation consists
in using

\begin{equation}
4\pi \left\langle R_{j}^{(\pi )}(r)R_{j^{\prime }}^{(\nu )}(r)\right\rangle
=N_{s}\;\rho \left( r\right)  \label{surf}
\end{equation}
where $\left\langle {}\right\rangle $ means the average over the $N_{s}$
participant surface nucleons, and is $\;\rho \left( r\right) $ the nuclear
density. To simplify, we also use $\ell =\ell ^{\prime }=0$, so that only
the spin of the nucleons are considered in the following developments. This
will limit $I^{\prime }=1$, and $I=0$. Thus, the relevant matrix element is

\begin{equation}
<\phi _{1/2}^{(\pi )}\,||\,[j_{0}(qr)\,\sigma \otimes Y_{0}]_{1}\,||\,\phi
_{1/2}^{(\nu )}>=-{1\over 2} \sqrt{\frac{3}{\pi }}{\cal F}_{0}\left( q\right) \;.
\label{m0}
\end{equation}

The Woods-Saxon distribution with central density $\rho _{0}$, radius $%
R_{0}, $ and diffusiveness $a,$ gives a good description of the densities of
the nuclei involved in our calculation. However, this distribution is very
well described by the convolution of a hard sphere and an Yukawa function 
\cite{DN76}. In this case, ${\cal F}_{0}\left( q\right) $, can be calculated
analytically \cite{spencer}

\begin{equation}
{\cal F}_{0}\left( q\right) =\frac{4\pi \rho _{0}}{q^{3}}\left[ \sin \left(
qR_{0}\right) -qR_{0}\cos \left( qR_{0}\right) \right] \left[ \frac{1}{%
1+q^{2}a_{Y}^{2}}\right]  \label{F0}
\end{equation}

Figure 1 compares ${\cal F}_{0}\left( q\right) $ obtained with the numerical
integration with the Woods-Saxon density distribution for $Al$ ($R_{0}=3.07$
fm , $a=0.519$ fm), $Cu$ ($R_{0}=4.163$ fm , $a=0.606$ fm), $Sn$ ($%
R_{0}=5.412$ fm , $a=0.560$ fm), $Au$ ($R_{0}=$ 6.43 fm , $a=0.$ 541 fm) ,
and $Pb$ ($R_{0}=6.62$ fm , $a=0.546$ fm) (to simplify the figure, we did
not plot the curves for $Au$ and $Cu$). In all cases we used for the Yukawa
function parameter in eq. (\ref{F0}) $a_{Y}=0.7$ fm. We see that the
agreement is very good. For carbon, one can use a Gaussian density, with the
Gaussian parameter $a=\sqrt{\frac{2}{3}\left\langle r^{2}\right\rangle _{C}}%
=2.018$ fm$^{2}$. In this case,

\begin{equation}
{\cal F}_{0}\left( q\right) =\pi ^{3/2}\rho _{0}a^{5/2}\exp \left(
-q^{2}a^{2}/4\right) \;.  \label{F02}
\end{equation}

A further simplification can be obtained for the factor $T(b)=\exp \left(
-2\,Im\;\chi (b)\right) $ in equation (\ref{prob}). We use the ''t$\rho \rho 
$'' approximation \cite{HRB91}, which gives 
\begin{equation}
T(b)=\exp \left\{ -\sigma _{NN}\,\int_{-\infty }^{\infty }dz\int \,\rho
_{P}\,({\bf r}\,)\rho _{T}\left( {\bf R}+{\bf r}\right) d^{3}r\right\} 
\label{tb}
\end{equation}
\noindent with $\,{\bf R}=({\bf b},z)\,$. $\sigma _{NN}\,$ is the
nucleon-nucleon cross section at the corresponding bombarding energy, and $%
\,\rho _{P(T)}\,$ is the projectile (target) matter density
distribution.\thinspace $T(b$)\thinspace is known as the transparency
function. At 158 GeV/nucleon we use $\sigma _{NN}$\thinspace $=52$ mb.

The simplest parameterization for the nuclear matter densities are Gaussian functions.
Assuming

\begin{equation}
\rho _{P(T)}\;(r)=\rho _{P(T)}\;(0)\,\exp \left\{ -r^{2}/\alpha
_{P(T)}^{2}\right\}   \label{rhopt}
\end{equation}
the integrals in eq.~(\ref{tb}) can be performed analytically. One gets 
\begin{equation}
T(b)=\exp \left\{ -\frac{\pi ^{2}\,\sigma _{NN}\;\rho _{T}(0)\;\rho
_{P}(0)\;\alpha _{T}^{3}\alpha _{P}^{3}}{\left( \alpha _{T}^{2}+\alpha
_{P}^{2}\right) } \;\exp \left[ -\frac{b^{2}}{\left( \alpha
_{T}^{2}+\alpha _{P}^{2}\right) }\right]\right\} \, .   \label{tbgauss}
\end{equation}

As observed by Karol \cite{Ka75}, for nuclei which densities well described
by Woods-Saxon distributions, $T(b)$ is very little dependent on the lower
values of $\,b\,$ and consequently on the values of $\,\rho _{P(T)}\;(r)\,$
for small $\,r^{\prime }s\,$. Only the surface form of the density is
relevant. Thus one can fit the surface part of the densities by Gaussian functions
and use the eq.~(\ref{tbgauss})$\,$ with the appropriate fitting parameters $%
\,\rho _{P}(0),$ $\rho _{T}(0),$ $a_{T}\,$ and $\,a_{P}\,$. If the density
distributions is given by a Fermi, or Woods-Saxon, function 
\begin{equation}
\rho (r)=\rho _{0}\{1+\exp [(r-R)/(t/4.4)]\}^{-1}  \label{rhokarol}
\end{equation}
Karol \cite{Ka75} has shown that $T(b)$ can be well reproduced with Gaussian
fits for the nuclear densities, if the parameters in the Gaussian
distributions (\ref{rhopt}) are given by 
\begin{equation}
\alpha ^{2}=\frac{4Rt+t^{2}}{k},\;\;\;\;\;\;\;\;\;\;\;\;\;\;\;\rho (0)=\frac{%
1}{2}\,\rho _{0}\,e^{R^{2}/\alpha ^{2}}\;,  \label{a2}
\end{equation}
where

\begin{equation}
\rho _{0}=\frac{3A}{4\pi R^{3}\left[ 1+\left( \pi
^{2}t^{2}/19.36R^{2}\right) \right] }\;,\;\;\;\;\;\;\;\;\;\;\;k=4(\ln
5)=6.43775...  \label{rho0}
\end{equation}
and $\,A\,$ is the mass number.

In figure 2 we compare the Karol transparency functions (open circles) with
the ones obtained by a numerical integration of the integral (\ref{tb}) with
Woods-Saxon densities (solid circles). The agreement is excellent. Thus,
the calculation simplifies enormously with the use of this approximation.

In eq. (\ref{m1}), only the transverse part of ${\bf q}$ is needed. Using 
\begin{eqnarray}
Y_{\ell m}(\hat{{\bf q}}_{t}) &=&(-1)^{(\ell +m)/2}\,\biggl( {\frac{2\ell +1%
}{4\pi }}\biggr)^{1/2}\,{\frac{\bigl[ (\ell -m)!\,(\ell +m)!\bigr]^{1/2}}{%
(\ell -m)!!\,(\ell +m)!!}}\,e^{im\phi }\,,\quad {\rm if}\quad \ell +m={\rm %
even} \nonumber \\
&=&0\,,\qquad {\rm otherwise}\,, \label{Yell}
\end{eqnarray}
and 
\begin{equation}
\int_{0}^{2\pi }\,e^{i(m-\nu )\phi }\,d\phi =2\pi \,\delta _{m,\nu }
\end{equation}
in the matrix elements (\ref{(B.5)}-\ref{(B.8)}) it is straightforward to
show that (\ref{m2}) becomes 
\[
M({\bf m},\,\nu ,\,b)={\cal C}_{0}(m_{P},\,m_{P}^{\prime
},m_{T},\,m_{T}^{\prime }\,)F_{0}(b)+{\cal C}_{\upsilon
}(m_{P},\,m_{P}^{\prime },m_{T},\,m_{T}^{\prime }\,)G_{\nu }(b),
\]
where 
\begin{eqnarray}
&&{\cal C}_{0}(m_{P},\,m_{P}^{\prime },m_{T},\,m_{T}^{\prime }\,)=\frac{%
3 }{2}\sum_{\mu }\,({{\frac{1}{2}}}1m_{P}\mu |{{\frac{1}{2}}}%
m_{P}^{\prime })({{\frac{1}{2}}}1m_{T}\mu |{{\frac{1}{2}}}m_{T}^{\prime }) 
\nonumber \\
&&{\cal C}_{\nu }(m_{P},\,m_{P}^{\prime },m_{T},\,m_{T}^{\prime }\,)=\frac{%
3}{4}\sum_{\mu \mu ^{\prime }}\,({{\frac{1}{2}}}1m_{P}\mu |{{\frac{1}{2}}%
}m_{P}^{\prime })({{\frac{1}{2}}}1m_{T}\mu ^{\prime }|{{\frac{1}{2}}}%
m_{T}^{\prime })\delta _{\mu -\mu ^{\prime }-\nu }
\end{eqnarray}
and

\begin{equation}
F_{0}(b)=\int_{0}^{\infty }dq\;qJ_{0}(qb)w(q){\cal F}_{0}^2\left( q\right)
,\;\;\;\;\;\;G_{\nu }(b)=\int_{0}^{\infty }dq\;qJ_{\nu }(qb)v(q){\cal F}%
_{0}^2\left( q\right) \;.
\end{equation}

In figure 3 we plot the charge-pickup probabilities for Pb
(158 GeV/nucleon) + X (target), with X = C, Al, Cu, Sn, Au and Pb. The
exchange probability is peaked at grazing impact parameters: at low impact
parameters the strong absorption makes the probability small, whereas at
large impact parameters it is small because of the short-range of the
exchange potentials. The value of the exchange probability at the peak is
about $10^{-5}$ for Pb + Pb. As in the case studied in ref. \cite{Ber93} the
process is dominated by $\pi $-exchange, with only a small fraction, of the
order of 10\%, or less, originating from the $\rho $ exchange channel.

The probabilities are divided by the square of the number of participating
nucleons. As shown, it increases with the target mass number. The cross
sections are found out to be approximately constant for bombarding energies
above 10 GeV/nucleon. The reason is simple, since from eq. (\ref{prob}) we
see that, apart from the factor $\left( 1/v\right) ^{2}\sim \left(
1/c\right) ^{2}$, the only energy dependence comes from the total
nucleon-nucleon cross section in the absorption factor $T(b)$. This is
approximately constant at relativistic energies.

In figure 4 we plot the total charge-pickup cross sections 
in Pb (158 GeV/nucleon) + X (target), with X = C, Al, Cu, Sn, Au and Pb. 
The dashed line is a guide to the eyes. The solid line is explained in
next section. The
cross section is also divided by the square of the number of participating
nucleons. It increases steadily with the target mass number.

In the table below we give the values of $\sigma /N_{s}^{2}$ (in mb) for the
studied reactions.

\bigskip
\begin{center}
\begin{tabular}{cc}
\hline\hline
Pb (158 GeV/nuc) + X \ \ \ \ \ \ \ & $\sigma /N_{s}^{2}$ (in mb) \\ \hline
C & 0.0106 \\ 
Al & 0.0394 \\ 
Cu & 0.105 \\ 
Sn & 0.253 \\ 
Au & 0.572 \\ 
Pb & 0.587 \\ \hline\hline
\end{tabular}
\end{center}

{\bf Table I} - {\it Total charge-pickup cross sections (per participant
nucleon) for Pb (158 GeV/nucleon) + X (target), with X = C, Al, Cu, Sn, Au and
Pb. }\bigskip \bigskip

\section{\bf Proportionality to the Gamow-Teller transition strengths}

It is clear that the approximation (\ref{surf}) is very rough and that the
assumption of $\ell =0$ nucleons at the surface is not realistic, specially
because the surface nucleons are certainly not in an s-wave state. However,
these approximations have been useful to extract the main features of the
reaction mechanism in relativistic $\pi +\rho $ exchange. Some of these
features are very useful for further theoretical developments. It is clear
from figures 2 and 3 that only nucleons very close to the surface will
contribute to the process. Thus, one can simplify these approximations by
replacing the nucleon coordinates in eq. (\ref{(B.5)}) by their surface
positions, ${\bf R}_{P}$ and ${\bf R}_{T}$, respectively. In this
case eq. (\ref{(B.5)}) reduces to the very compact expression

\begin{equation}
{\mathcal M}=\left[ w(q)+v(q)\,\right] \,\,M_{GT}\left( P\longrightarrow
P^{\prime }\right) \;M_{GT}\left( T\longrightarrow T^{\prime }\right) \;e^{i%
\mathbf{q.R}_{P}}\;e^{i\mathbf{q.R}_{T}}  \label{mnew}
\end{equation}
where we also used eq. (\ref{Yell}).
The
Gamow-Teller matrix elements $M_{GT}$ are expectation values of the $\sigma
\tau $ operators. I.e.,

\begin{equation}
M_{GT}\left( A\longrightarrow A^{\prime }\right) =\int d^{3}r\;\rho _{\sigma
\tau }\left( \mathbf{r}\right) =\langle A^{\prime }\left| \sigma \tau
\right| A\rangle \;.  \label{MGT}
\end{equation}

Inserting eq. (\ref{mnew}) \ into eqs. (\ref{m1}) and (\ref{prob}), and using the
integral

\begin{equation}
\int d\phi \;\exp \left( -i\nu \phi \right) \;\exp (-iqx\cos \phi )=2\pi
J_{\nu }\left( qx\right) \;,
\end{equation}
we get 
\begin{eqnarray}
{\mathcal P}(b) &=&\biggl({\frac{1}{4\pi ^{2}\hbar v}}\biggr)^{2}\;\exp %
\bigl\{-2\,Im\;\chi (b)\bigr\}\,\;  
B_{GT}\left( P\longrightarrow P^{\prime }\right) \,B_{GT}\left(
T\longrightarrow T^{\prime }\right) \; 
\nonumber \\
&\times& \sum_{\nu }\left| H\left( \nu
,\;b\right) \right| ^{2}\,,  \label{probf}
\end{eqnarray}
where

\begin{equation}
B_{GT}\left( A\longrightarrow A^{\prime }\right) =\left| M_{GT}\left(
A\longrightarrow A^{\prime }\right) \,\right| ^{2}=\left| \langle A^{\prime
}\left| \sigma \tau \right| A\rangle \right| ^{2}  \label{BGT}
\end{equation}
is the Gamow-Teller transition strength of nucleus $A$. A sum over final
spins and average over initial spins is implicit. The function $H\left( \nu ,\;b\right) $ is given by

\begin{equation}
H\left( \nu ,\;b\right) =2\pi \int_{0}^{\infty }dq\;q\;\left[
w(q)+v(q)\,\right] \;J_{\nu }\left( qb\right) \;J_{\nu }\left[ q\left(
R_{P}+R_{T}\right) \right] \;.  \label{H}
\end{equation}
This function is peaked at $b=$ $R_{P}+R_{T}$ and its width is determined by
the range of the pion-exchange interaction, i.e., $\Delta b\sim \;1\;$%
fm.

The expression (\ref{probf}) shows the proportionality between the charge
exchange probabilities and the Gamow-Teller transition strengths in
relativistic heavy ion collisions. A similar relationship was obtained by
Taddeucci et al. \cite{Ta87} for (p, n) reactions at $0$ degrees for 
proton energies of $\sim $ 100 MeV. 
For heavy-ion reactions at $0$ degrees and bombarding  
energies of $\sim $ 100 MeV, a similar relationship was also obtained  
by Osterfeld et al. \cite{Ost92}.
The validity of such a proportionality
depends on the factorization of the cross sections into two terms, one
depending on the nuclear sizes and absorption, and the other on the nuclear
structure. The solid line in figure 4 displays the cross sections for
charge-pickup reactions using the proportionality expression (\ref{probf}).
The results have been normalized to yield the same magnitude as $\sigma
/N_{s}^{2}$ for the reaction Pb (158 GeV/nucleon) + Sn. One sees that the
agreement is excellent, showing that the A-dependence of the process is
solely due to geometry factors (nuclear transparency and the range
of the one-boson-exchange interaction). The oversimplified model of the last section
is thus unnecessary, but it was useful to show that one can use the
proportionality expression (\ref{probf}) to access precious information on the
beta-decay strengths from charge-pickup reactions with relativistic heavy
ions.

\section{\bf Conclusions}

We described the charge exchange in relativistic heavy ion reactions in
terms of $\pi $ and $\rho $ exchange and the eikonal approximation. 
We applied the formalism to the
calculation of charge-pickup probabilities and cross sections for
Pb (158 GeV/nucleon) + X (target), with X = C, Al, Cu, Sn, Au and
Pb.

Our model yields probabilities and cross sections which are dependent on the
number of participating nucleons at the nuclear surface.  The cross sections for the process are not small, as seen from
figure 4. Assuming, e.g., that the number of surface nucleons which can
contribute to the process is of order of 10 for large systems, one gets
cross sections of order of 50 mb, or more.

The calculation is useful to support the proportionality of the 
measured cross sections with the Gamow-Teller matrix transition strengths.
These are useful nuclear structure information. However, the experiments
would have to be able to distinguish these "elastic" charge-exchange 
events from more complicated backgrounds, e.g., charge-exchange with pion
production.

\newpage

\section{\noindent {\bf Acknowledgments}}

\medskip

The authors have benefited from many suggestions and fruitful discussions
with C. Scheidenberger and K. Suemmerer.

This work was partially supported by the Brazilian funding agencies FAPERJ,
CNPq, FUJB/UFRJ, and by MCT/FINEP/CNPQ(PRONEX) under contract No.
41.96.0886.00.

\section{{\bf Figure Captions}.}

\begin{enumerate}
\item  The from factor ${\cal F}_{0}\left( q\right) $ of eq. (\ref{F0}),
compared to the one obtained by the numerical integration using Woods-Saxon
density distributions for $Al$, $Cu$, $Sn$, $Au$, and $Pb$ (to simplify the
figure, we did not plot the curves for $Au$ and $Cu$).

\item  Comparison of the Karol transparency functions (open circles) with
the ones obtained by a numerical integration of the integral \ref{tb} with
Woods-Saxon densities (solid circles).

\item  Charge-pickup probabilities for Pb (158 GeV/nucleon)
+ X (target), with X = C, Al, Cu, Sn, Au and Pb.

\item  Total charge-pickup cross sections for Pb (158
GeV/nucleon) + X (target), with X = C, Al, Cu, Sn, Au and Pb. The dashed curve is a guide to the eyes. The solid curve represents a calculation of the cross sections divided by the 
Gamow-Teller transition strenghts of the nuclei and normalized to the $\sigma/N^2$ cross
section for Pb + Sn.

\end{enumerate}

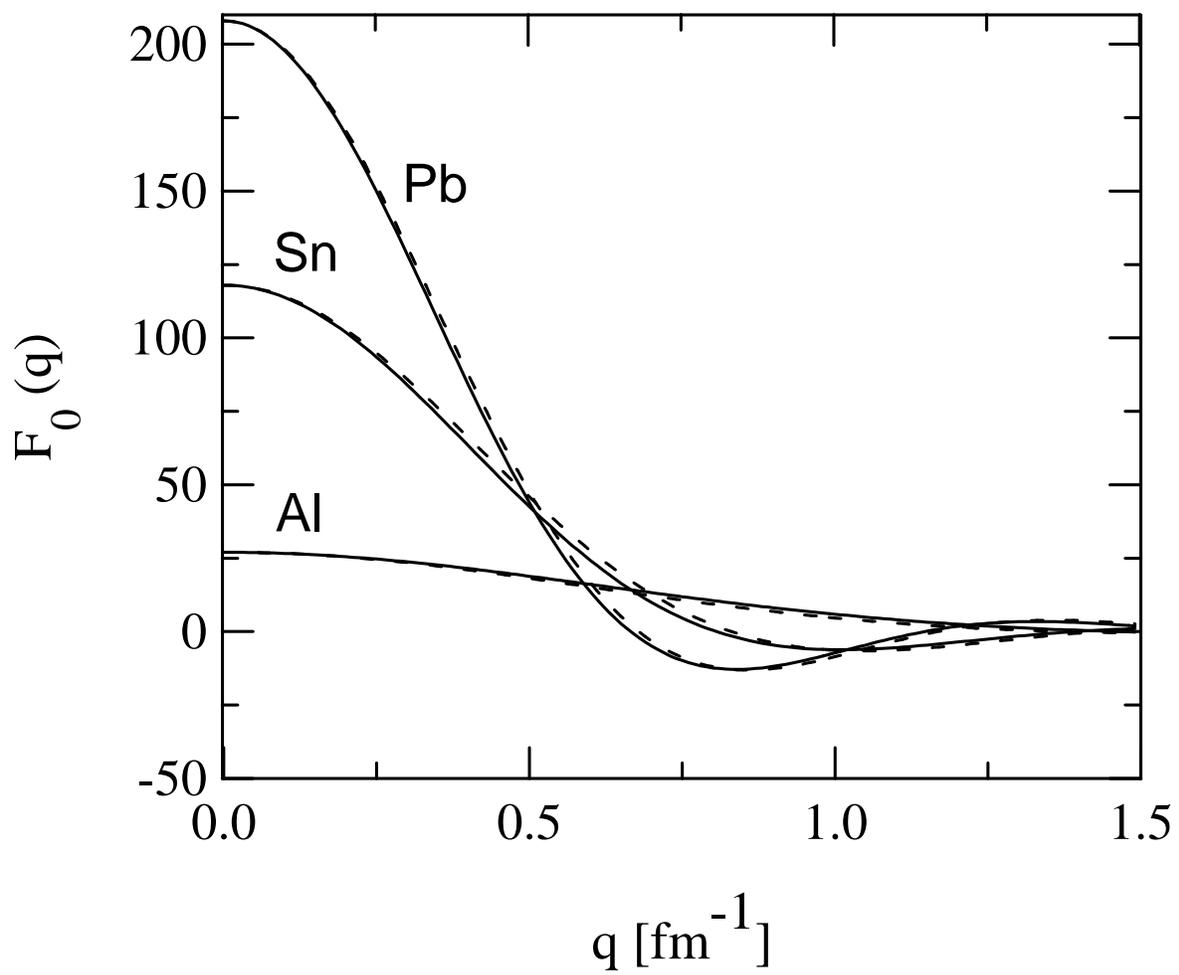
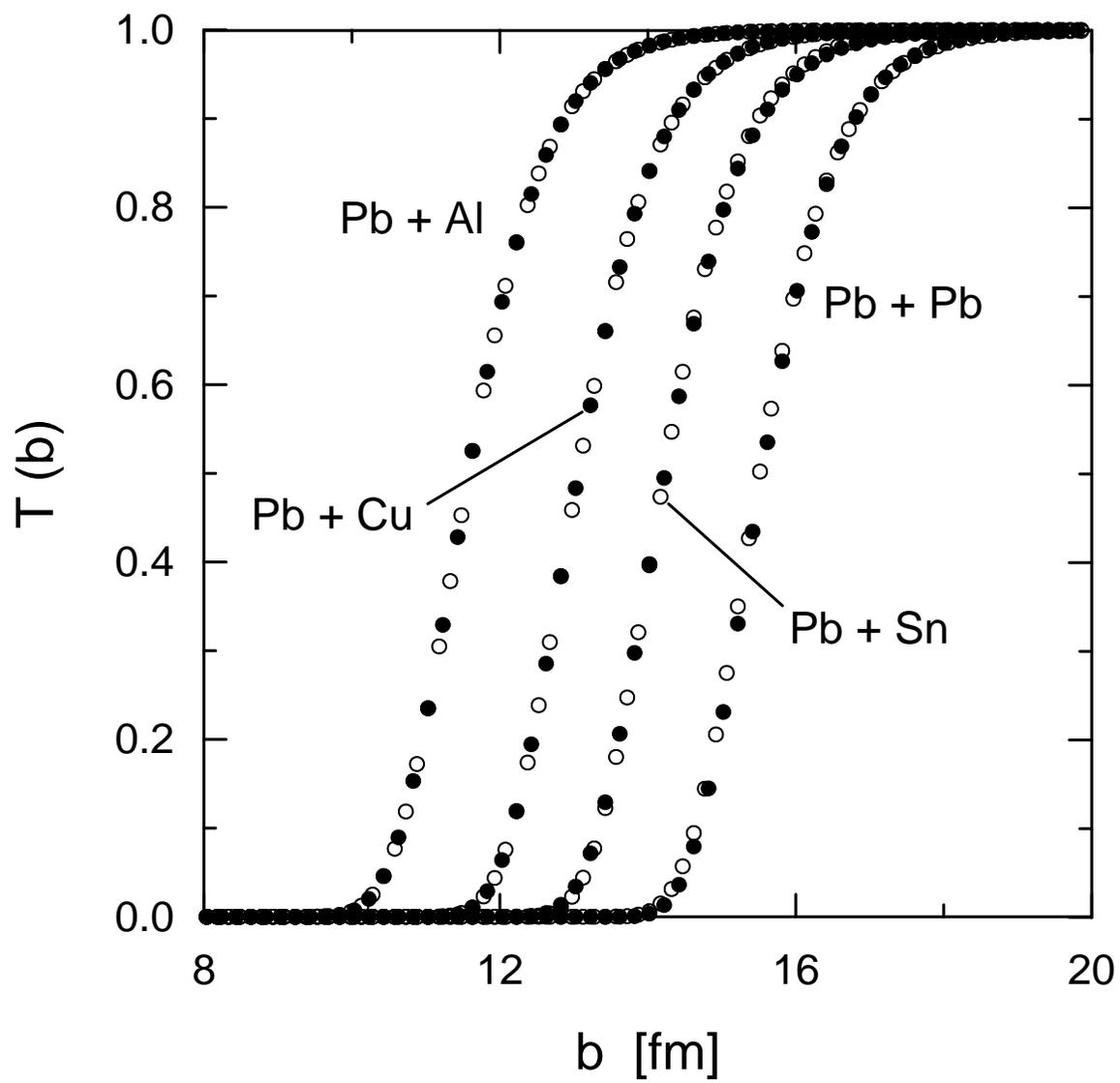
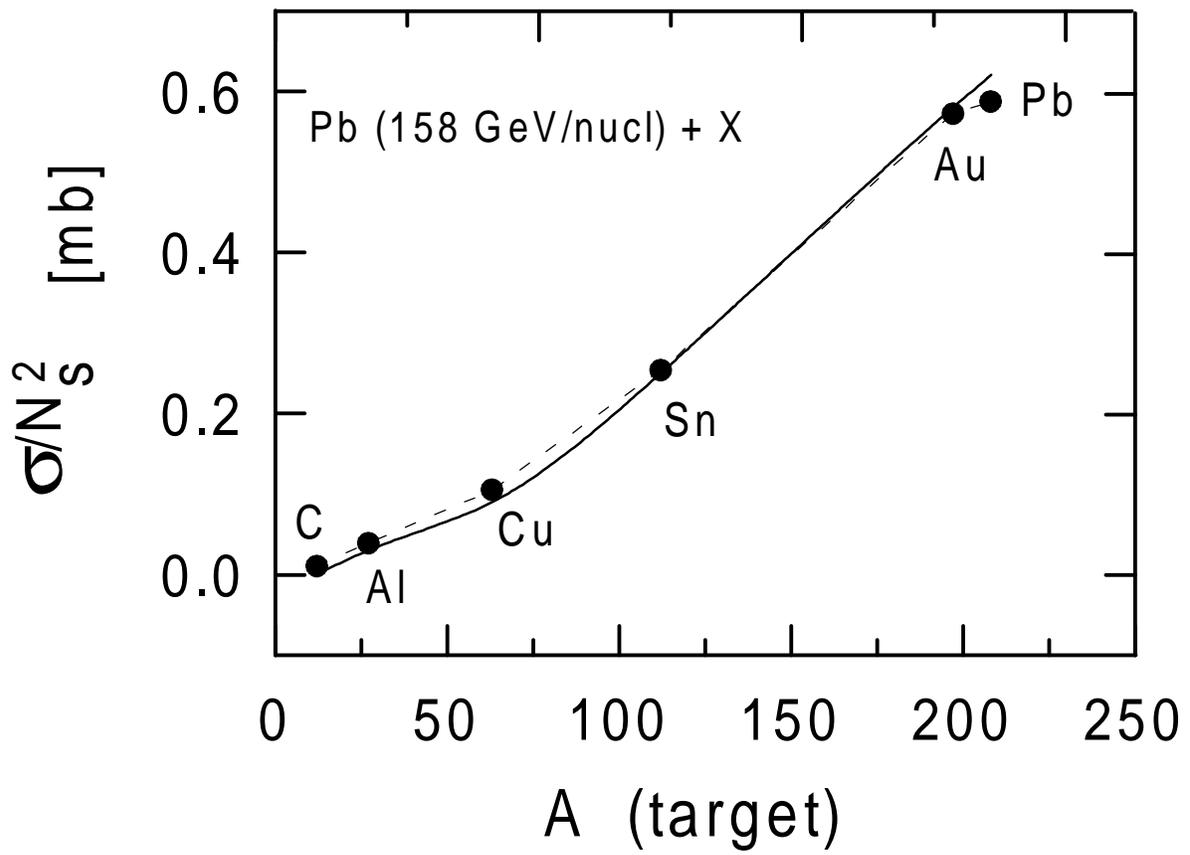
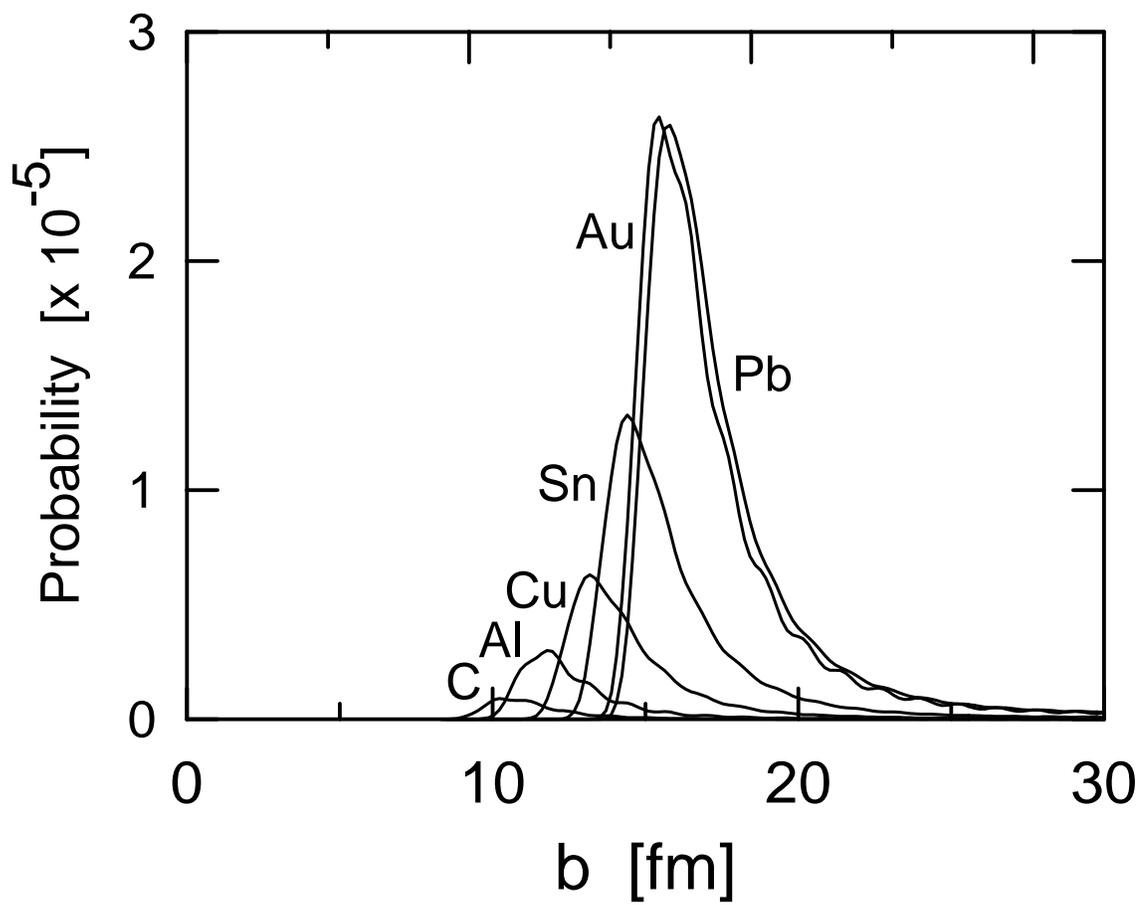
\end{document}